\documentclass[twocolumn,10pt]{IEEEtran}
\DeclareUnicodeCharacter{3000}{ }
\DeclareUnicodeCharacter{03D5}{  }
\usepackage{textcomp}
\usepackage[bottom=0.78in,top=0.72in,left=0.625in,right=0.625in]{geometry}

\usepackage{epstopdf}
\usepackage{multirow}
\usepackage[cmex10]{amsmath}
\usepackage{amssymb}
\usepackage{amsthm}
\usepackage{array}

\usepackage{url}
\usepackage{nomencl}
\usepackage{cite}
\usepackage{mathtools}
\usepackage{makeidx}
\usepackage{ifthen}
\usepackage{subfigure}
\usepackage{caption}
\usepackage{color}
\usepackage{xcolor}
\usepackage{pdflscape}
\usepackage{bm,upgreek}
\usepackage{float}
\usepackage{algorithm}
\usepackage{algorithmic}
\usepackage{tabularx} 
\usepackage{booktabs} 

\makenomenclature

\renewcommand{\nomgroup}[1]{%
\ifthenelse{\equal{#1}{I}}{\item[\textbf{Indices}]}{%
\ifthenelse{\equal{#1}{A}}{\item[\textbf{Abbreviations}]}{%
\ifthenelse{\equal{#1}{V}}{\item[\textbf{Variables}]}{%
\ifthenelse{\equal{#1}{P}}{\item[\textbf{Parameters and Constants}]}{%
}
}
}
}
}

\newcommand{\beq}{\begin{equation}}
\newcommand{\eeq}{\end{equation}}
\newcommand{\beqn}{\begin{eqnarray}}
\newcommand{\eeqn}{\end{eqnarray}}
\newcommand{\beqno}{\begin{eqnarray*}}
\newcommand{\eeqno}{\end{eqnarray*}}
\newcommand{\bma}{\begin{displaymath}}
\newcommand{\ema}{\end{displaymath}}
\newcommand{\bnu}{\begin{enumerate}}
\newcommand{\enu}{\end{enumerate}}
\newcommand{\bce}{\begin{center}}
\newcommand{\ece}{\end{center}}
\newcommand{\btb}{\begin{tabular}}
\newcommand{\etb}{\end{tabular}}

\begin{document}

\title{Reactive Power-Aware Data Center–Grid Coordination for Voltage Regulation in Unbalanced Three-Phase Distribution Systems}

\title{A Bilevel Framework for Data Center-Grid Coordination with DLMPs in Unbalanced Three-Phase Distribution Systems}

\author{\IEEEauthorblockN{Arash~Baharvandi,~\IEEEmembership{Student Member,~IEEE} and~Duong~Tung~Nguyen,~\IEEEmembership{Member,~IEEE}} 
\thanks{The authors are with the School of Electrical, Computer and Energy Engineering, Arizona State University, Tempe, AZ, United States. Email: \textit\{abaharv1,~duongnt\}@asu.edu.}}

 \maketitle
\thispagestyle{empty} 

\begin{abstract}
This paper proposes a grid-aware coordination framework between data centers and distribution grids using a DLMP-based bilevel optimization model. The data center aggregator (DCA) determines active power demand in response to distribution locational marginal prices (DLMPs), while the distribution system operator (DSO) solves a network-constrained optimal power flow problem to determine DLMPs in an unbalanced three-phase system. The model incorporates both active and reactive power consumption of data centers to evaluate their impacts on voltage regulation and phase imbalance. To mitigate adverse network effects, two operating cases are analyzed: without reactive power compensation and with static var generator (SVG)-based compensation. The proposed approach is validated on the IEEE 37-bus unbalanced distribution test system. Simulation results show that DLMP-based coordination captures economically efficient data center operation, and phase- and location-dependent network conditions, while SVG-based compensation improves voltage profiles and reduces phase unbalance.
\end{abstract}


\allowdisplaybreaks[4]

\section{Introduction}
The rapid growth of cloud computing, social networks, big data, and Internet of Things (IoT) has driven large-scale data center expansion worldwide. As capacity and energy demand rise, the data center rack market is projected to grow from \$4.6 billion in 2025 to \$9.1 billion by 2034 \cite{imarcgroup}. Data centers already accounted for about 3\% of global electricity consumption in 2016 \cite{masanet2020recalibrating}, and with emerging artificial intelligence workloads, their energy use may reach up to 21\% of global demand by 2030, creating major challenges for efficient and reliable power system operation.

Several studies have examined the role of data centers in supporting power system operation \cite{liu2024optimal,ellaicnc24,ding2025multi,wang2020flexible}. Delay-tolerant workloads can serve as flexible resources to improve renewable energy integration and operational efficiency through optimal scheduling \cite{liu2024optimal}. Geo-distributed data centers have been coordinated to reduce operational costs by considering electricity prices, renewable generation, and energy storage \cite{ellaicnc24}. They can also participate in local flexibility markets to support coordinated regional energy management and balance flexibility supply and demand \cite{ding2025multi}. In addition, data centers have been modeled as flexible loads for wind power integration and curtailment mitigation under uncertainty using stochastic scheduling \cite{wang2020flexible}.

Several studies have used bilevel optimization \cite{bilevel22} to model data center--grid coordination \cite{wang2015proactive,wang2014stackelberg,wu2023incentivizing,yang2023distribution}. These frameworks commonly design grid-side pricing strategies to guide workload distribution across geographically dispersed data centers, improving load balancing and reducing operational costs \cite{wang2015proactive}. Stackelberg game models capture hierarchical decisions, where the grid sets pricing signals and data centers optimize workload allocation for economic performance \cite{wang2014stackelberg}. Data center flexibility has also been coordinated through incentive-based bilevel models to improve system efficiency and renewable energy integration \cite{wu2023incentivizing}. More recently, distribution locational marginal price (DLMP)-based bilevel frameworks have enabled data centers to schedule workloads and energy storage in response to locational prices while satisfying network constraints \cite{yang2023distribution}. \textit{However, none of these studies consider unbalanced three-phase distribution systems within a bilevel framework for coordinating data centers and power grids.}

Some studies have incorporated reliability and stability considerations when integrating data centers into power grid operation \cite{gyang2025dynamic,ali2017ancillary,aksanli2015minimizing}. Dynamic models show that data centers are sensitive to grid disturbances, such as load variations and faults, highlighting the need for accurate representation to ensure reliable operation \cite{gyang2025dynamic}. Data centers have also been used as computational resources for ancillary services, with optimal job scheduling and service-level agreements improving efficiency and reliability \cite{ali2017ancillary}. Moreover, variations in data center power consumption can affect grid stability and cause voltage instability, motivating coordinated workload control strategies that maintain performance requirements \cite{aksanli2015minimizing}. \textit{However, existing studies do not explicitly capture the impact of data center reactive power consumption on voltage.}

Some studies have investigated voltage unbalance in three-phase distribution networks \cite{wang2017phase,liu2022bi,pinthurat2023simultaneous}. Data center integration can influence phase imbalance, while coordination with distributed energy resources can improve power quality, reliability, and operational efficiency \cite{wang2017phase}. Bilevel optimization has also been used to mitigate three-phase unbalance in low-voltage systems through optimal phase switch placement while considering investment and operational costs \cite{liu2022bi}. In addition, coordinated control of distributed resources, such as photovoltaic inverter reactive power, can simultaneously regulate voltage and reduce unbalance \cite{pinthurat2023simultaneous}. \textit{Existing studies do not explicitly capture the impact of data center reactive power consumption on voltage regulation in unbalanced three-phase distribution systems within a bilevel framework.}

To address the aforementioned gaps, this paper proposes a DLMP-based bilevel optimization framework for coordinated operation between data centers and distribution systems. In the proposed approach, the DCA determines its operational strategy in response to pricing signals, while the distribution system operator (DSO) is responsible for the economic dispatch of distributed generations (DGs) while respecting operational and technical constraints, including voltage limits and power flow conditions. The framework explicitly captures the impact of both active and reactive power consumption of data centers on voltage regulation and phase unbalance in unbalanced three-phase distribution systems. In addition, reactive power support is incorporated to mitigate adverse network effects and enhance system performance. The main contributions of this work are summarized as follows:
\begin{itemize}
\item \textbf{\textit{Modeling}}: A DLMP-based bilevel coordination model is developed to quantify the interaction between data center load scheduling and distribution-grid operation in unbalanced three-phase networks. The model explicitly incorporates data center reactive power consumption and evaluates its impact on voltage regulation and phase imbalance.

\item \textbf{\textit{Solution Approach}}: The proposed bilevel problem is reformulated into a tractable single-level optimization problem using Karush–Kuhn–Tucker (KKT) conditions. Standard linearization techniques, including McCormick formulations and binary expansion, are employed to enable efficient computation while preserving the economic and operational interactions between the DCA and DSO.

\item \textbf{\textit{Numerical Results}}: Simulation studies on the IEEE 37-bus unbalanced three-phase test feeder demonstrate how DLMPs influence data center load allocation and how SVG-based reactive power compensation mitigates voltage drops and phase-dependent network stress caused by data center loading.
\end{itemize}
The remainder of this paper is organized as follows. Section~\ref{model} describes the system model, and Section~\ref{formulation} presents the problem formulation. Simulation results are provided in Section~\ref{Results}, followed by conclusions in Section~\ref{Conclusion}.

\begin{table}[h]
\caption{NOTATIONS}
\centering
\small 
\begin{tabularx}{\columnwidth}{|l|X|} 
\hline
\textbf{Notation} & \textbf{Meaning} \\ \hline
\multicolumn{2}{|c|}{\textbf{Sets and Indices}} \\ \hline
$n,\mathcal{N}$               & Index and set of buses \\ \hline
$\phi,\Phi$ & Index and set of phases \\ \hline
$g,\mathcal{G},\mathcal{G}^{n}$        & Index and set of distributed generators (DGs), and set of DGs on bus $n$\\ \hline
$l$,$\mathcal{L}$ & Index and set of lines\\
\hline
$\mathcal{L}^{out}, \mathcal{L}^{in}$ & Sets of outgoing and incoming lines\\ \hline 
$t,\mathcal{T}$      & Indices and set of periods\\ \hline
$i,\mathcal{I},\mathcal{N}^{i}$      & Indices and set of data center facility, and set of buses with facilities \\ \hline
\multicolumn{2}{|c|}{\textbf{Parameters}} \\ \hline
$C_{g,\phi,t}$& DG cost function\\ \hline
$p^{{d}}_{n,\phi,t},q^{{d}}_{n,\phi,t}$ & Active/reactive load at bus $n$, phase $\phi$ and time $t$\\ \hline
$r_{\phi,\phi^0},x_{\phi,\phi^0}$ & Self and mutual impedance between phases \\ \hline 
$\underline{p}_g,\overline{p}_g$ & Minimum/maximum active capacity for DG $g$
\\ \hline 
$\underline{q}_g,\overline{q}_g$ & Minimum/maximum reactive capacity for DG $g$
\\ \hline 
$\overline{f}^{p}_l,\overline{f}^{q}_l$ & Capacity for active and reactive power on line $l$
\\ \hline 
$\underline{v}_n,\overline{v}_n$ & Minimum and maximum squared voltage at bus $n$\\ \hline
$P^{id}_{i}, P^{max}_{i}$  & Server idle and maximum active power \\ \hline
$\Delta_t$&Duration of period $t$\\ \hline 
$PUE_{i}$  & Power usage efficiency for facility $i$ \\ \hline
$F^{r}_{i}$&Server service rate  \\ \hline $W^{max}_{i}$ &Maximum number of servers at facility $i$  \\ \hline $DT$ & The maximum delay time \\ \hline $F^{fe}_{t}$ & Front-end server workload forecast \\ \hline $\tilde{Q}^{s}_{i}$ & SVG capacity \\ \hline
\multicolumn{2}{|c|}{\textbf{Variables}} \\ \hline
$p_{g,\phi,t},q_{g,\phi,t}$ & Active and reactive power for DG $g$ at phase $\phi$ and period $t$ \\ \hline 
$f^{p}_{l,\phi,t},f^{q}_{l,\phi,t}$& Active and reactive power flows through line $l$ at phase $\phi$ and period $t$  \\ \hline 
$v_{n,\phi,t}$& Squared voltage at bus $n$, phase $\phi$ and period $t$  \\ \hline
$P^{dc}_{i,t},Q^{dc}_{i,t}$ & Active and reactive power of facility $i$ at period $t$  \\ \hline
$\lambda_{n,\phi,t}^p$& DLMP at bus $n$, phase $\phi$, and period $t$ \\ \hline
$ W_{i,t}$& Number of active servers for facility $i$ at period $t$  \\ \hline
$F_{i,t}$ &Total arriving workloads at facility $i$ and period $t$  \\ \hline $Q^{s}_{i,t}$ & Reactive power of SVG at facility $i$ and period $t$ \\ \hline 
\end{tabularx}

\label{notation}
\end{table}

\vspace{-0.2cm}
\section{System Model }
\label{model}
In this paper, we consider a data center aggregator (DCA) managing geographically distributed data centers connected to an unbalanced three-phase distribution network. The DCA minimizes operational costs by scheduling active power consumption in response to distribution locational marginal prices (DLMPs), while the distribution system operator (DSO) performs economic dispatch subject to network, operational, and technical constraints, including power flow and voltage limits. The proposed framework captures the hierarchical coordination between the DCA and DSO, where the DCA responds to pricing signals and the DSO determines DLMPs based on network conditions.
The data center power consumption is explicitly modeled because of its impact on voltage profiles and phase unbalance, and reactive power support devices, such as static var generators (SVGs), are incorporated to mitigate these effects. This bidirectional coupling between data center operation and network performance motivates a bilevel optimization model. In practice, data center electricity tariffs are typically based on fixed, demand-based, or time-of-use pricing rather than phase-specific DLMPs due to regulatory, metering, and billing limitations. Thus, DLMPs are not assumed to be directly passed to individual data centers as retail prices; instead, they are internalized by the DCA as coordination signals for workload allocation and power consumption across facilities while respecting distribution-network constraints.

\begin{figure}[h!]
\vspace{- 0.2cm}
	\centering		\includegraphics[width=0.5\textwidth,height=0.15\textheight]{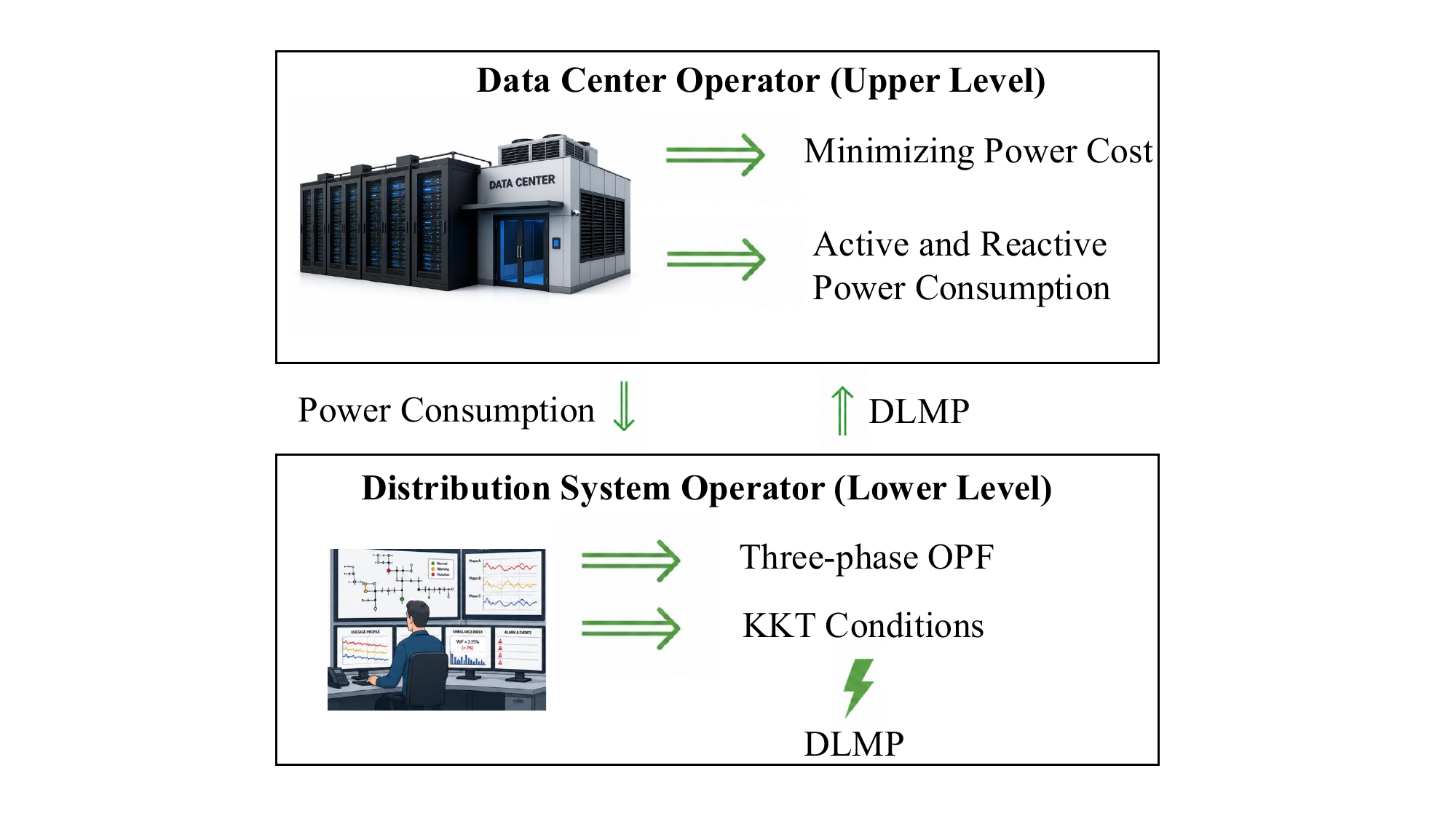}
			\caption{System Model}
	\label{fig:model}
    \vspace{- 0.2cm}
\end{figure}

The proposed model focuses on the electrical interaction between data centers and unbalanced distribution grids. Other flexibility costs, including thermal cycling, cooling dynamics, server wear, workload migration overhead, and infrastructure-level degradation, are not explicitly modeled. These hidden costs for data center operators can be incorporated in future work. Fig.~\ref{fig:model} illustrates the bilevel interaction between the DCA and the DSO.

Let $\mathcal{N}$ denote the set of buses in the distribution network, where $n$ represents the bus index. The set of phases is denoted by $\Phi$, with $\phi$ indicating the phase index. Let $\mathcal{G}$ be the set of distributed generators (DGs), and $\mathcal{G}^{n}$ denote the subset of DGs connected to bus $n$, where $g$ is the DG index. The set of distribution lines is represented by $\mathcal{L}$, with $l$ as the line index, and $\mathcal{L}^{out}$ and $\mathcal{L}^{in}$ denote the sets of outgoing and incoming lines, respectively. The time horizon is indexed by $t \in \mathcal{T}$, where $\mathcal{T}$ is the set of periods. Moreover, $\mathcal{I}$ denotes the set of data center facilities, and $\mathcal{N}^{i}$ represents the set of buses hosting data center facilities, where $i$ is the facility index.

The active and reactive power cost of DG $g$ at phase $\phi$ and time $t$ is represented by $C_{g,\phi,t}$. The active and reactive demands at bus $n$, phase $\phi$, and time $t$ are denoted by $p^{d}_{n,\phi,t}$ and $q^{d}_{n,\phi,t}$, respectively. The self and mutual impedances between phases are given by $r_{\phi,\phi^0}$ and $x_{\phi,\phi^0}$. The minimum and maximum active and reactive capacities for DGs are denoted by $\underline{p}_g$ and $\overline{p}_g$, $\underline{q}_g$ and $\overline{q}_g$, respectively, while $\overline{f}^{p}_l$ and $\overline{f}^{q}_l$ represent the active and reactive power flow limits of line $l$. The squared voltage magnitude limits at bus $n$ are specified by $\underline{v}_n$ and $\overline{v}_n$. For each data center facility $i$, $P^{id}_{i}$ and $P^{max}_{i}$ denote the idle and maximum server active power consumption, $PUE_{i}$ is the power usage efficiency, $F^{r}_{i}$ is the server service rate, $W^{max}_{i}$ is the maximum number of active servers, $DT$ is the maximum delay time, $F^{fe}_{t}$ is the front-end workload forecast, and $\tilde{Q}^{s}_{i}$ is the SVG capacity. The duration of each period is given by $\Delta_t$.

The decision variables include the DG active and reactive power outputs $p_{g,\phi,t}$ and $q_{g,\phi,t}$, the active and reactive line flows $f^{p}_{l,\phi,t}$ and $f^{q}_{l,\phi,t}$, and the squared voltage magnitude $v_{n,\phi,t}$ at each bus, phase, and time period. The data center active and reactive power consumptions are represented by $P^{dc}_{i,t}$ and $Q^{dc}_{i,t}$, respectively, while $\lambda_{n,\phi,t}^{p}$ denotes the DLMP at bus $n$, phase $\phi$, and time $t$. In addition, $W_{i,t}$ is the number of active servers at facility $i$, $F_{i,t}$ is the total arriving workload, and $Q^{s}_{i,t}$ is the reactive power injected by the SVG at facility $i$ and time $t$.
\section{Problem Formulation}
\label{formulation}
This section formulates coordinated data center--grid operation as a bilevel optimization problem. The upper level models a data center aggregator (DCA), which allocates computational workloads and manages local energy resources across geographically distributed facilities to minimize electricity procurement cost. The lower level represents the distribution system operator (DSO), which determines distribution locational marginal prices (DLMPs) by solving a network-constrained optimal power flow (OPF) problem subject to distribution-network constraints. The two levels are coupled through DLMPs, which link data center decisions with grid operation.
\subsection{Upper-Level Model: Data Center Aggregator}

At the upper level, the DCA acts as the leader and coordinates multiple data center facilities. The electrical demand of each facility includes both IT-related consumption (e.g., servers, networking equipment) and auxiliary consumption such as cooling and lighting. The relationship between total power consumption and IT load is characterized by the power usage efficiency (PUE) \cite{jiang2021optimal}.

The DCA aims to minimize its total electricity procurement cost based on nodal prices provided by the lower-level distribution 
operator. Upper-level formulation is given as follows:
\vspace{-0.5cm}
\begin{subequations} \label{eq:upper_level}
\begin{align}
& \min_{\substack{\small \boldsymbol{P}^{dc},\boldsymbol{W},\\\boldsymbol{F},\boldsymbol{Q^{S}}}} \sum_{i \in \mathcal{N}^i} \sum_{\phi \in \Phi} \sum_{t\in \mathcal{T}}   \lambda_{i,\phi,t}^p P^{dc}_{i,t} \Delta_t 
\label{dca-obj}\\
& \textit{s.t:} ~~ 
P^{dc}_{i,t} = \left[P^{id}_{i} + \left(PUE_{i} - 1\right) P^{max}_{i}\right] W_{i,t} 
\nonumber \\&+ \frac{P^{max}_{i} - P^{id}_{i}}{F^{r}_{i}} F_{i,t}, 
~~ \forall i,t \label{eq:power_model} \\
&0 \leq  W_{i,t} \leq W^{max}_{i}, 
 ~~~\forall i,t \label{eq:server_limit} \\
& \frac{1}{DT} \leq F^{r}_{i}  W_{i,t}- F_{i,t} ,~~~ \forall i,t \label{eq:qos} \\
& F_{i,t} \geq 0, 
~~~\forall i,t \label{eq:workload_pos} \\& \sum_{i} F_{i,t} \geq F^{fe}_{t}, 
~~~\forall t \label{eq:workload_balance} \\
& -{\tilde Q}^{s}_{i} \leq Q^{s}_{i,t} \leq  0 
~~~ \forall i,t. \label{eq:svc} 
\end{align}
\end{subequations}

The objective function (\ref{dca-obj}) minimizes the total electricity procurement cost of the data center aggregator, where $\lambda_{i,\phi,t}$ represents the DLMP at node $i$, phase $\phi$ and time $t$, obtained from the lower-level problem. Constraints~(\ref{eq:power_model}) define the total active power consumption of data center facility $i$ at time $t$. The terms $P_i^{id}W_{i,t}$ and $(PUE_i-1)P_i^{max}W_{i,t}$ capture idle server power and non-information-technology auxiliary power, such as cooling and power delivery overhead, respectively. The term $\frac{P_i^{max}-P_i^{id}}{F_i^r}F_{i,t}$ represents workload-dependent information-technology power. Thus, $P_{i,t}^{dc}$ includes both baseline and workload-dependent power consumption. Constraints (\ref{eq:server_limit}) limit the available number of active servers. Constraints (\ref{eq:qos}) represent the quality-of-service (QoS) requirement for a non-zero workload, where the waiting time is modeled based on M/M/1 queueing theory \cite{wan2017joint}. The parameter $DT$ represents the maximum allowable delay and defines the QoS limit of the data center operation. A smaller value of $DT$ enforces a stricter QoS requirement and reduces the flexibility available for DLMP-based load adjustment, while a larger value increases flexibility at the expense of looser latency performance. Constraints (\ref{eq:workload_pos}) ensure the non-negativity of the arriving workload, while constraints (\ref{eq:workload_balance}) guarantee that the total incoming workload demand is satisfied across all data center facilities.
Also, constraints (\ref{eq:svc}) represent the operational limits of SVG. The forecast workload and quality-of-service (QoS) requirement are assumed feasible within available data center capacity; otherwise, additional capacity, workload shedding, or QoS relaxation would be required.
\subsection{Lower-Level Problem: Distribution System Operator}
\label{sec:lower_level}

At the lower level, the DSO optimizes network operation by solving a network-constrained economic dispatch problem. The goal is to minimize the total generation cost while satisfying power flow and operational constraints, given the aggregate data center demand provided by the upper-level problem.
\begin{subequations}
\label{eq:lower_level}
\begin{align}
\min_{\substack{\boldsymbol{p,q},\, \\ \boldsymbol{f^{p},\,f^{q}},\boldsymbol{v}}}
& \sum_{g \in \mathcal{G}} \sum_{\phi \in \Phi} \sum_{t \in \mathcal{T}}   C_{g,\phi,t}(p_{g,\phi,t})
\label{eq:lower_obj} \\[4pt]
\textit{s.t.} \quad
& \sum_{g \in \mathcal{G}^{n}} p_{g,\phi,t}
- \sum_{l \in \mathcal{L}^{{out}}} f^{p}_{l,\phi,t}
+ \sum_{l \in \mathcal{L}^{{in}}} f^{p}_{l,\phi,t}
= p^{{d}}_{n,\phi,t}  \nonumber \\&+ \sum_{i \in \mathcal{N}^{i}}P
^{dc}_{i,t} ~~~ \forall n,\phi,t
\label{eq:power_balance_p} ~~~ (\lambda_{n,\phi,t}^p) \\[4pt]
& \sum_{g \in \mathcal{G}^{n}} q_{g,\phi,t}
- \sum_{l \in \mathcal{L}^{{out}}} f^{q}_{l,\phi,t}
+ \sum_{l \in \mathcal{L}^{{in}}} f^{q}_{l,\phi,t}
= q^{{d}}_{n,\phi,t}  \nonumber \\&+\sum_{i \in \mathcal{N}^{i}}Q
^{dc}_{i,t}+ \sum_{i \in \mathcal{N}^{i}}Q
^{s}_{i,t} ~~~ \forall n,\phi,t
\label{eq:power_balance_q} ~~~ (\lambda_{n,\phi,t}^q) \\[4pt]
& v_{m(l),\phi,t} - v_{n(l),\phi,t}
\!\!= \!\!\ \sum_{\phi^0 \in \Phi}(h_{\phi,\phi^0,l} f^{p}_{l,\phi^0,t} \nonumber\\ &+ b_{\phi,\phi^0,l} f^{q}_{l,\phi^0,t}), 
~~~ \forall l  \in \mathcal{L},t,\phi,
\label{eq:voltage_drop} ~~~ (\beta_{l,\phi,t}) \\
& \!\mathbf{H}_{\phi\phi^0} \!\!=\!\!\!
\begin{aligned}
\begin{bmatrix}
-2r_{aa} & \!\!r_{ab}\!-\!\sqrt{3}x_{ab} & \!r_{ac}\!+\!\sqrt{3}x_{ac} \\
\!r_{ba}\!+\!\sqrt{3}x_{ba} & -2r_{bb} & r_{bc}\!-\!\sqrt{3}x_{bc} \\
\!r_{ca}\!-\!\sqrt{3}x_{ca} & \!\!r_{cb}\!+\!\sqrt{3}x_{cb} & -2r_{cc}
\end{bmatrix}_{\!l} \!\!\!\!
\end{aligned}
\label{eq:H} \\
& \mathbf{B}_{\phi\phi^0} \!\!=\!\!\!
\begin{aligned}
\begin{bmatrix}
-2x_{aa} &\!\! x_{ab}\!+\!\sqrt{3}r_{ab} & \!x_{ac}\!-\!\sqrt{3}r_{ac} \\
\!x_{ba}\!-\!\sqrt{3}r_{ba} & -2x_{bb} & x_{bc}\!+\!\sqrt{3}r_{bc} \\
\!x_{ca}\!+\!\sqrt{3}r_{ca} & \!\!x_{cb}\!-\!\sqrt{3}r_{cb} & -2x_{cc}
\end{bmatrix}_{l}\!\!\!\!
\end{aligned}
\label{eq:B}\\
& \underline{p}_g \le p_{g,\phi,t} \le \overline{p}_g,
~~~ \forall g,\phi,t~~~(\alpha^{p-}_{g,\phi,t},\alpha^{p+}_{g,\phi,t})
\label{eq:gen_p_limits} \\[4pt]
& \underline{q}_g \le q_{g,\phi,t} \le \overline{q}_g,
~~~ \forall g,\phi,t~~~(\alpha^{q-}_{g,\phi,t},\alpha^{q+}_{g,\phi,t})
\label{eq:gen_q_limits} \\[4pt]
& -\overline{f}^{p}_l \le f^{p}_{l,\phi,t} \le \overline{f}^{p}_l,
~~~ \forall l,\phi,t~~~(\sigma^{p-}_{l,\phi,t},\sigma^{p+}_{l,\phi,t})
\label{eq:flow_p_limits} \\[4pt]
& -\overline{f}^{q}_l \le f^{q}_{l,\phi,t} \le \overline{f}^{q}_l,
~~~ \forall l,\phi,t~~~(\sigma^{q-}_{l,\phi,t},\sigma^{q+}_{l,\phi,t})
\label{eq:flow_q_limits} \\[4pt]
& \underline{v}_n \le v_{n,\phi,t} \le \overline{v}_n,
~~~ \forall n,\phi,t~~~(\tau^{-}_{n,\phi,t},\tau^{+}_{n,\phi,t})
\label{eq:voltage_limits}
\end{align}
\end{subequations}
The DSO minimizes the total generation cost of DGs in the network, as defined in (\ref{eq:lower_obj}). For modeling simplicity, the upstream grid is more expensive than existing DGs, and grid does not import power from upstream.
Constraints (\ref{eq:power_balance_p})–(\ref{eq:power_balance_q}) enforce active and reactive power balance at each node and phase for every time period. The reactive power of the data center is modeled as $Q^{dc}_{i,t}=P^{dc}_{i,t}tan(cos^{-1}(PF))$ , where PF denotes the power factor of the data center. Constraints (\ref{eq:voltage_drop}) describe the load flow equations for an unbalanced three-phase network, capturing voltage drops along distribution lines as functions of active and reactive power flows, while neglecting line losses \cite{li2025distributed}. $h_{\phi,\phi^0,l}$ and $b_{\phi,\phi^0,l}$ denote the elements of the matrices defined in (\ref{eq:H}) and (\ref{eq:B}), respectively.  It is noted that $a$, $b$, and $c$ denote phases in these matrices. Generator operating limits on active and reactive power are imposed through constraints (\ref{eq:gen_p_limits})–(\ref{eq:gen_q_limits}). Line flow limits for active and reactive power are enforced in constraints (\ref{eq:flow_p_limits})–(\ref{eq:flow_q_limits}), while bus voltage magnitude limits are maintained through (\ref{eq:voltage_limits}). The variables shown in parentheses after each constraint denote the corresponding dual variables. 
\vspace{-0.3cm}
\subsection{Solution Approach}
\label{thd:Solution Approach}
To transform the bilevel optimization problem into a single-level formulation, the Karush–Kuhn–Tucker (KKT) conditions of the lower-level problem are incorporated into the upper-level model. This reformulation is valid due to the convexity of the lower-level problem, which ensures that the KKT conditions are both necessary and sufficient for optimality. By embedding the stationarity, primal feasibility, dual feasibility, and complementary slackness conditions, the lower-level (follower) problem is implicitly represented within the upper-level (leader) problem. 
The KKT conditions of the lower-level problem are given as follows:
\begin{subequations}
\label{KKT}
\begin{align}
&\frac{\partial C_{g,\phi,t}}{\partial p_{g,\phi,t}}-\lambda^{p}_{n,\phi,t}+\alpha^{p+}_{g,\phi,t}-\alpha^{p-}_{g,\phi,t}=\!0,~\!\forall t,\phi, (g,n) \!\!\in \!\!\mathcal{G}^{n}
\label{kkt1}\\
&\lambda^{p}_{n,\phi,t}-\lambda^{p}_{m,\phi,t}+\sum_{\phi^0}h_{\phi^0,\phi,l}\beta_{l,\phi^0,t}+\sigma^{p+}_{l,\phi,t}-\sigma^{p-}_{l,\phi,t}\nonumber \\&=0,\forall l,\phi,t
\label{kkt2}\\
&\lambda^{q}_{n,\phi,t}-\lambda^{q}_{m,\phi,t}+\sum_{\phi^0}b_{\phi^0,\phi,l}\beta_{l,\phi^0,t}+\sigma^{q+}_{l,\phi,t}-\sigma^{q-}_{l,\phi,t}\nonumber \\&=0,\forall l,\phi,t
\label{kkt3}\\
&-\lambda^{q}_{n,\phi,t}+\alpha^{q+}_{g,\phi,t}-\alpha^{q-}_{g,\phi,t}=0,~~~\forall t,\phi,(g,n) \in \mathcal{G}^{n}
\label{kkt4}\\
&\sum_{l \in \mathcal{L}^{{out}}} \!\beta_{l,\phi,t}-\sum_{l \in \mathcal{L}^{in}} \!\beta_{l,\phi,t}-\tau^{-}_{n,\phi,t}+\tau^{+}_{n,\phi,t}=0,~\forall n,\phi,t
\label{kkt5}\\
&(\underline{p}_g- p_{g,\phi,t})\alpha^{p-}_{g,\phi,t}=0,~~~\forall g,\phi,t
\label{kkt6}\\
&(p_{g,\phi,t}-\overline{p}_g)\alpha^{p+}_{g,\phi,t}=0,~~~\forall g,\phi,t
\label{kkt7}\\
&(\underline{q}_g- q_{g,\phi,t})\alpha^{q-}_{g,\phi,t}=0,~~~\forall g,\phi,t
\label{kkt8}\\
&(q_{g,\phi,t}-\overline{q}_g)\alpha^{q+}_{g,\phi,t}=0,~~~\forall g,\phi,t
\label{kkt9}\\
&(-\overline{f}^{p}_l-f^{p}_{l,\phi,t})\sigma^{p-}_{l,\phi,t}=0,~~~\forall l,\phi,t
\label{kkt10}\\
&(-\overline{f}^{p}_l+f^{p}_{l,\phi,t})\sigma^{p+}_{l,\phi,t}=0,~~~\forall l,\phi,t
\label{kkt11}\\
&(-\overline{f}^{q}_l-f^{q}_{l,\phi,t})\sigma^{q-}_{l,\phi,t}=0,~~~\forall l,\phi,t
\label{kkt12}\\
&(-\overline{f}^{q}_l+f^{q}_{l,\phi,t})\sigma^{q+}_{l,\phi,t}=0,~~~\forall l,\phi,t
\label{kkt13}\\
&(\underline{v}_n-v_{n,\phi,t})\tau^{-}_{n,\phi,t}=0,~~~\forall n,\phi,t
\label{kkt14}\\
&(-\overline{v}_n+v_{n,\phi,t})\tau^{+}_{n,\phi,t}=0,~~~\forall n,\phi,t
\label{kkt15}\\
&\boldsymbol{\alpha^{p-}}, \!\boldsymbol{\alpha^{p+}}, \!\boldsymbol{\alpha^{q-}}, \!\boldsymbol{\alpha^{q+}}, \!\boldsymbol{\sigma^{p-}}, \!\boldsymbol{\sigma^{p+}}, \!\boldsymbol{\sigma^{q-}}, \!\boldsymbol{\sigma^{q+}}, \!\boldsymbol{\tau^{-}}, \!\boldsymbol{\tau^{+}} \!\!\geq \!0.
\label{kkt16}
\end{align}
\end{subequations}
Constraints (\ref{kkt1})–(\ref{kkt5}) correspond to the stationarity conditions of the Lagrangian. The complementary slackness conditions are enforced through (\ref{kkt6})–(\ref{kkt15}), while (\ref{kkt16}) ensures dual feasibility.
Based on these conditions, the bilevel problem can be reformulated as the following single-level optimization problem:
\begin{equation}
\label{single-level-reformulation}
\chi=: \{(\ref{eq:upper_level}, (\ref{eq:power_balance_p})–(\ref{eq:voltage_limits}), (\ref{KKT}) \}
\end{equation}

To linearize constraints (\ref{kkt6})–(\ref{kkt15}), the Big-M method \cite{risk15} is employed to reformulate the nonlinear complementary slackness conditions of the form $xy=0$ as follows:
\begin{subequations}
\begin{align}
& x \le M z, \quad y \le M (1 - z),\\
& x \ge 0,\; y \ge 0, \quad z \in \{0,1\}.
\end{align}
\end{subequations}
where $M$ is a big number. Additionally, the leader’s objective function (\ref{dca-obj}) is nonlinear due to the multiplication of upper- and lower-level variables. This nonlinearity can be addressed using binary expansion, as follows:
\begin{subequations}
\label{binary-expansion}
\begin{align}
 &
P^{dc}_{i,t}=\frac{P^{\max}_{i}}{2^{E}-1} \sum_{e \in \mathcal{E}} 2^{e} u_{i,t,e},~~~ \forall i \in \mathcal{N}^i,t \label{P_dc_expansion}
\\[4pt]
& b_{i,\phi,t,e} \le M\, u_{i,t,e}, 
~~~ \forall i,\phi,t,e \label{big-M1}\\
& b_{i,\phi,t,e} \ge -M\, u_{i,t,e}, 
~~~ \forall i,\phi,t,e  \label{big-M2}\\
& b_{i,\phi,t,e} \le \lambda^p_{i,\phi,t} + M(1 - u_{i,t,e}), 
~~~\forall i,\phi,t,e \label{big-M3} \\
& b_{i,\phi,t,e} \ge \lambda^p_{i,\phi,t} - M(1 - u_{i,t,e}), 
~~~\forall i,\phi,t,e \label{big-M4} \\[6pt]
& \min \sum_{i \in \mathcal{N}^{i}} \sum_{t \in \mathcal{T}} \sum_{\phi \in \Phi} \sum_{e \in \mathcal{E}} 
(\frac{P^{\max}_{i}}{2^{E}-1} )2^e  b_{i,\phi,t,e}  \Delta_
t \label{linear-obj}
\end{align}
\end{subequations}
The set of constraints in (\ref{binary-expansion}) presents a binary expansion-based linearization of the bilinear term involving the product of  $\lambda^p_{i,\phi,t}$ (DLMP) and the decision variable $P^{dc}_{i,t}$.
Constraints (\ref{P_dc_expansion}) represent $P^{dc}_{i,t}$ as a discretized value using binary expansion. The index $e \in \mathcal{E}=\{0,1,...,E-1\}$ denotes the binary expansion level (or digit) used to discretize the continuous variable. The set $\mathcal{E}$ contains all binary indices employed in the approximation, and its cardinality $E = |\mathcal{E}|$ represents the total number of binary variables used for the expansion. The binary variables $u_{i,t,e} \in \{0,1\}$ indicate the activation of each discretization level, while the factor $\frac{P^{\max}_{i}}{2^{E}-1}$ defines the resolution of the approximation.
To linearize the bilinear term $\lambda^p_{i,\phi,t} P^{dc}_{i,t}$, an auxiliary variable $b_{i,\phi,t,e}$ is introduced to represent the product $\lambda^p_{i,\phi,t} u_{i,t,e}$. Constraints (\ref{big-M1})--(\ref{big-M4}) enforce this relationship using a Big-$M$ formulation. Specifically, when $u_{i,t,e} = 1$, the variable $b_{i,\phi,t,e}$ is forced to equal $\lambda^p_{i,\phi,t}$, whereas when $u_{i,t,e} = 0$, it is restricted to zero.
Finally, the original nonlinear objective is reformulated in (\ref{linear-obj}) by substituting the bilinear term with its linearized equivalent. The coefficients $(\frac{P^{\max}_{i}}{2^{E}-1})2^e$ ensure consistency with the binary representation of $P^{dc}_{i,t}$, while $\Delta_t$ accounts for the time interval. Finally, the objective function in  (\ref{dca-obj}) should be replaced by (\ref{binary-expansion}). This transformation converts the problem into a mixed-integer quadratic programming (MIQP) formulation. The proposed formulation provides a tractable optimization-based characterization of the data center--grid coordination problem under linearized unbalanced power-flow assumptions and discretized linearization of bilinear terms. Thus, the obtained solution should be interpreted as a benchmark or lower-bound characterization of the coordinated operation rather than a complete real-time implementation framework. Practical deployment would require additional considerations, including AC power-flow validation, uncertainty in workloads and network conditions, communication delays, market rules, and detailed data center operational constraints.

\section{Simulation Results}
\label{Results}
\begin{figure}[h!]
\centering

\subfigure[DLMP at bus 703 ]{%
    \includegraphics[width=0.23\textwidth,height=0.11\textheight]{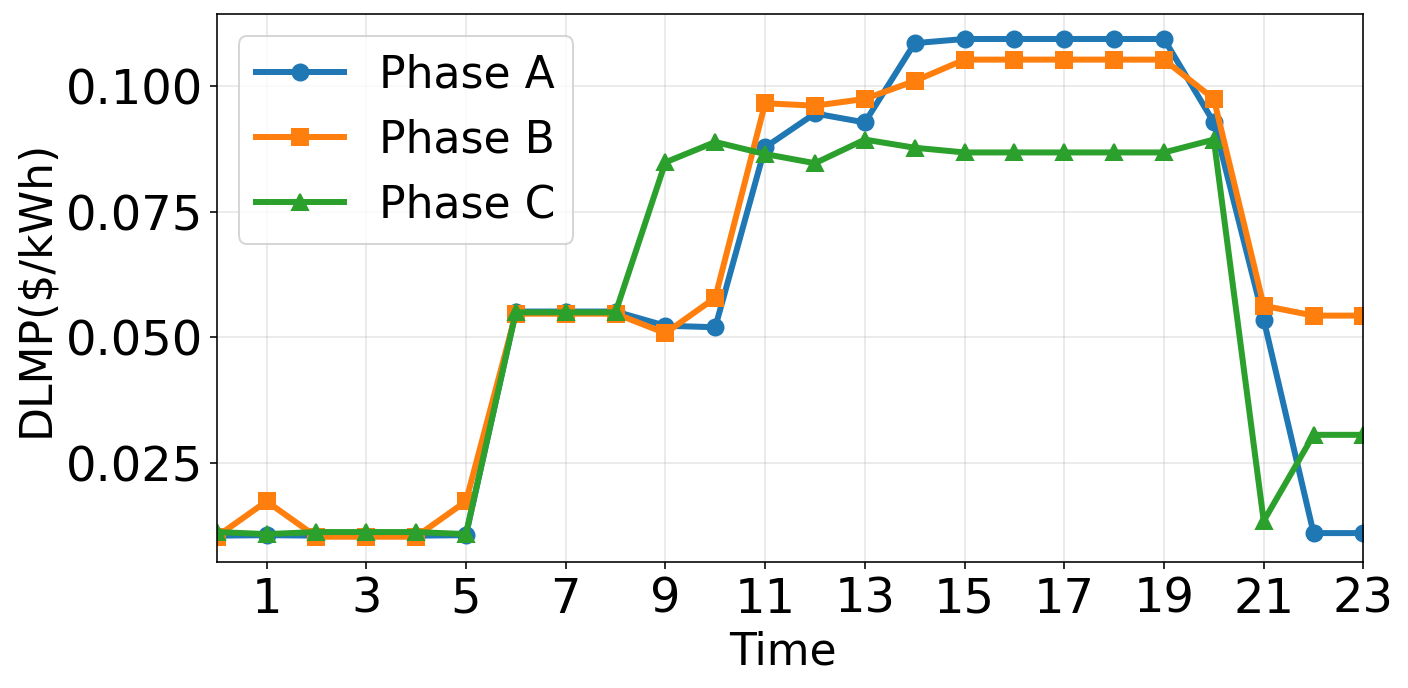}%
    \label{fig:lmpvtime703}%
}
\hspace*{-0.18cm}
\subfigure[DLMP at bus 733 ]{%
    \includegraphics[width=0.23\textwidth,height=0.11\textheight]{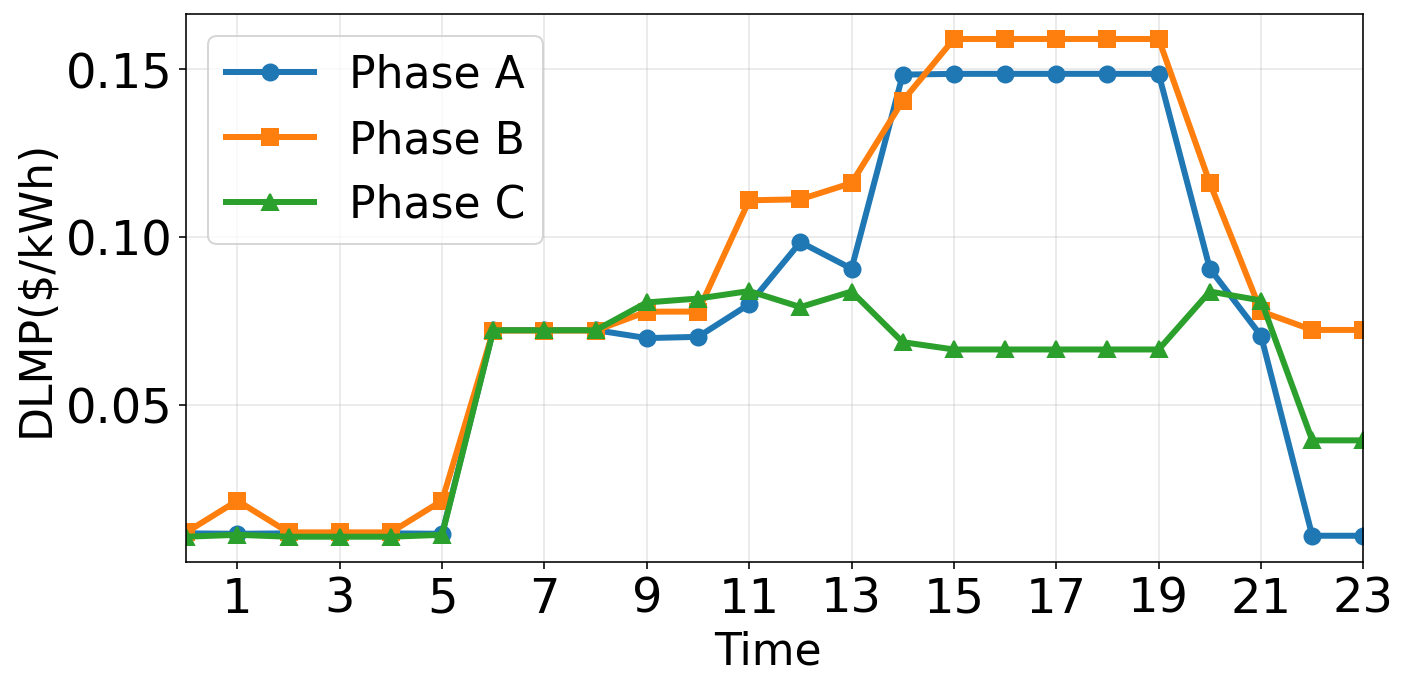}%
    \label{fig:lmpvtime733}%
}

\vspace{-0.2cm}

\subfigure[DLMP at bus 741 ]{%
    \includegraphics[width=0.23\textwidth,height=0.11\textheight]{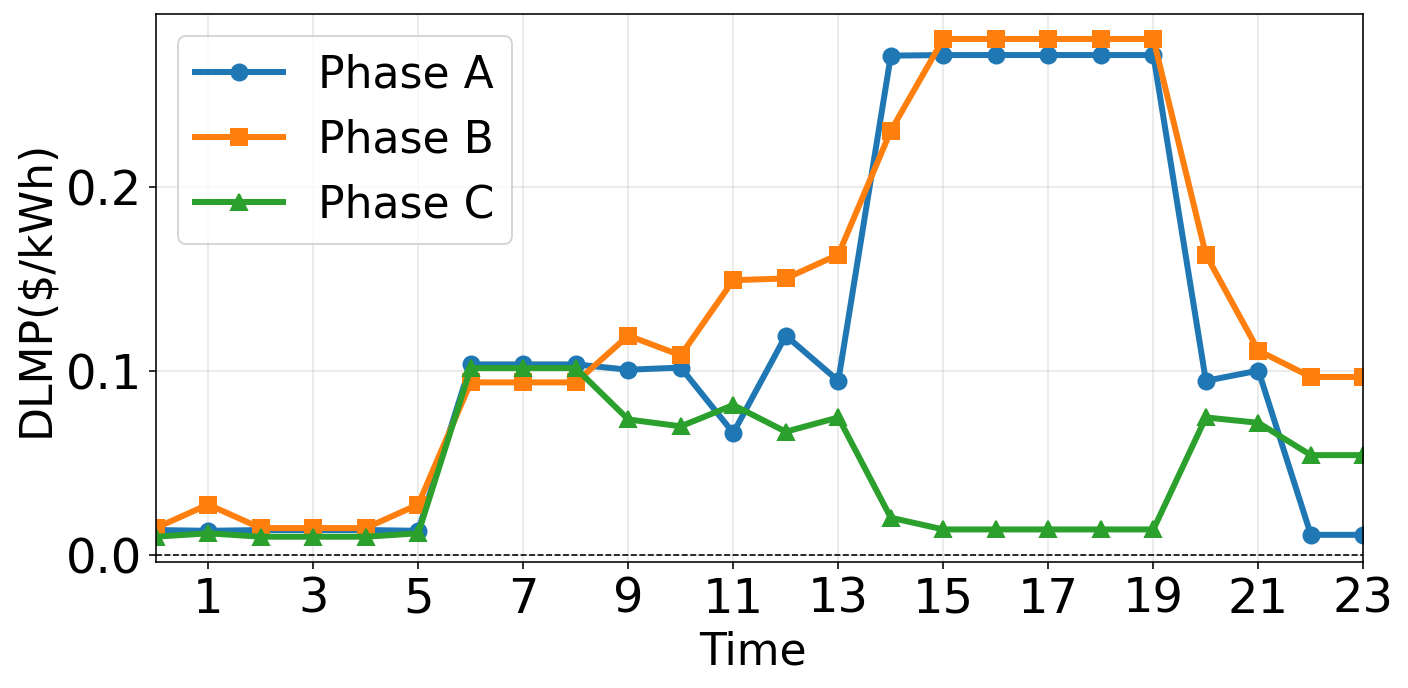}%
    \label{fig:lmpvtime741}%
}
\vspace{-0.4cm}
\caption{DLMP at data center nodes for all phases }
                \label{DLMPvtime}

\vspace{-0.2cm}
\end{figure}

\begin{figure}[h!]
    \centering
    \includegraphics[width=0.5\linewidth]{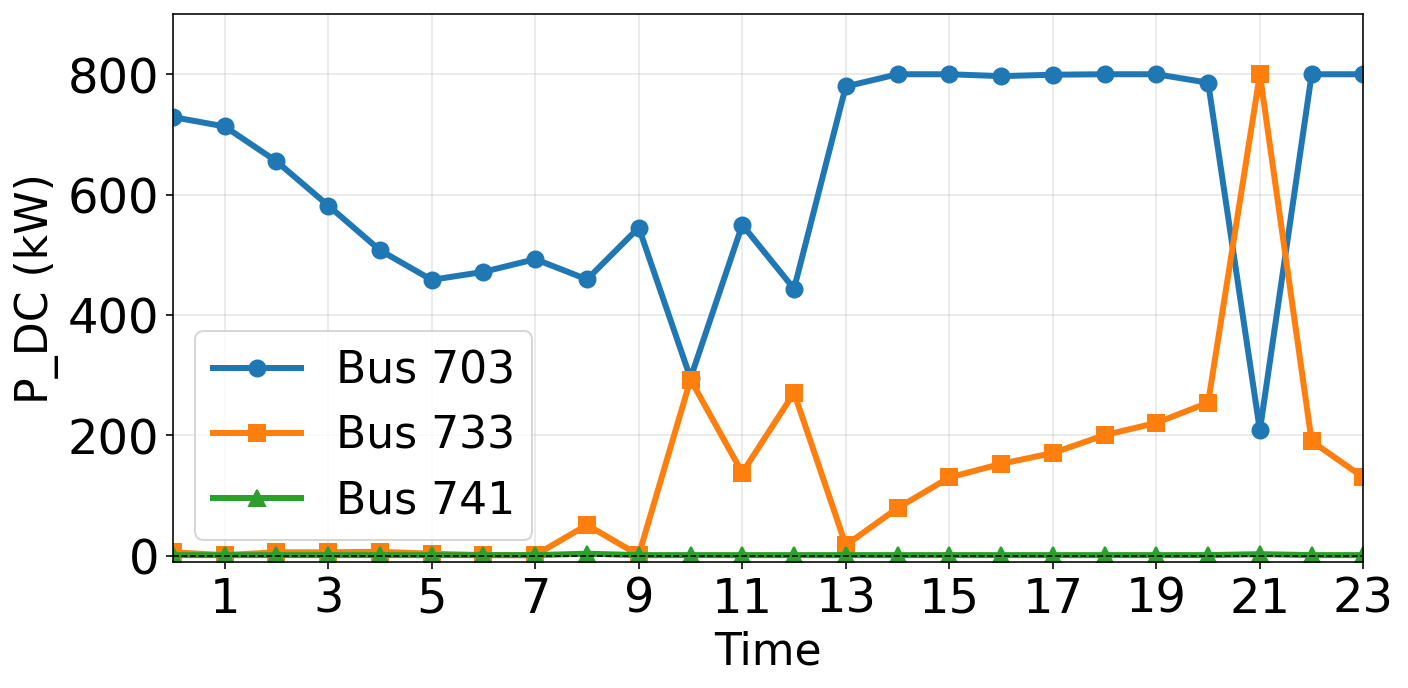}
    \vspace{-0.3cm}
    \caption{Data center power consumption on different nodes}
    \label{fig:P_dcvtime}
\end{figure}

\begin{figure}[h!]
\centering

\subfigure[Voltage at node 703 ]{%
    \includegraphics[width=0.23\textwidth,height=0.11\textheight]{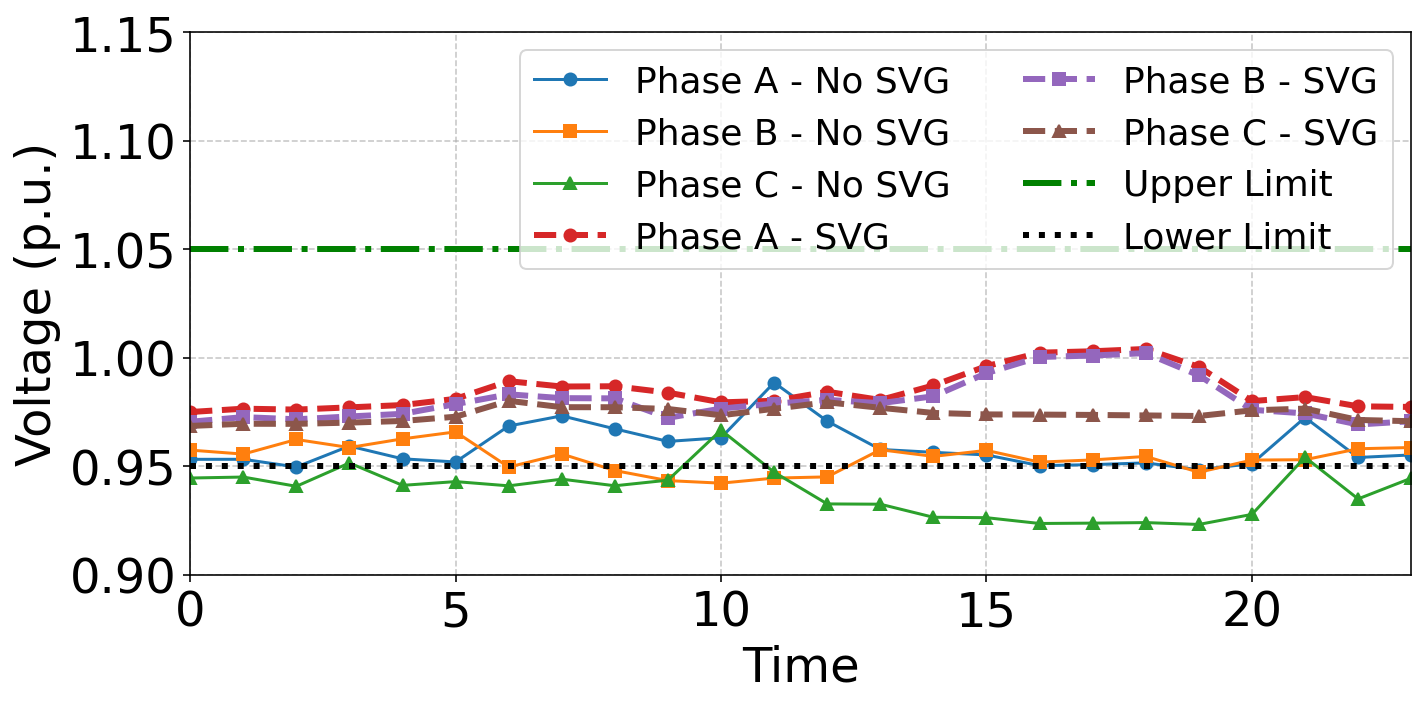}%
    \label{fig:voltagevtime703}%
}
\hspace*{-0.18cm}
\subfigure[Voltage at node 733 ]{%
    \includegraphics[width=0.23\textwidth,height=0.11\textheight]{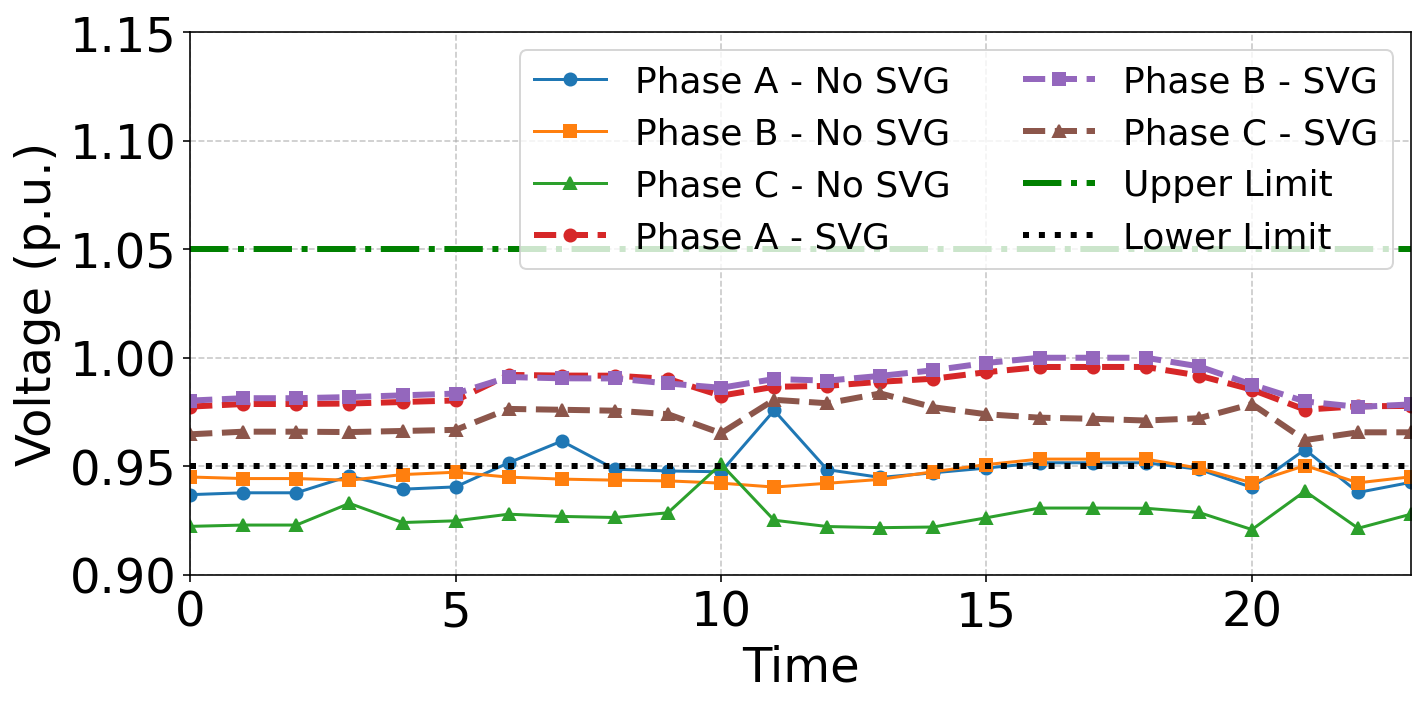}%
    \label{fig:voltagevtime733}%
}

\vspace{-0.2cm}

\subfigure[Voltage at node 741 ]{%
    \includegraphics[width=0.23\textwidth,height=0.11\textheight]{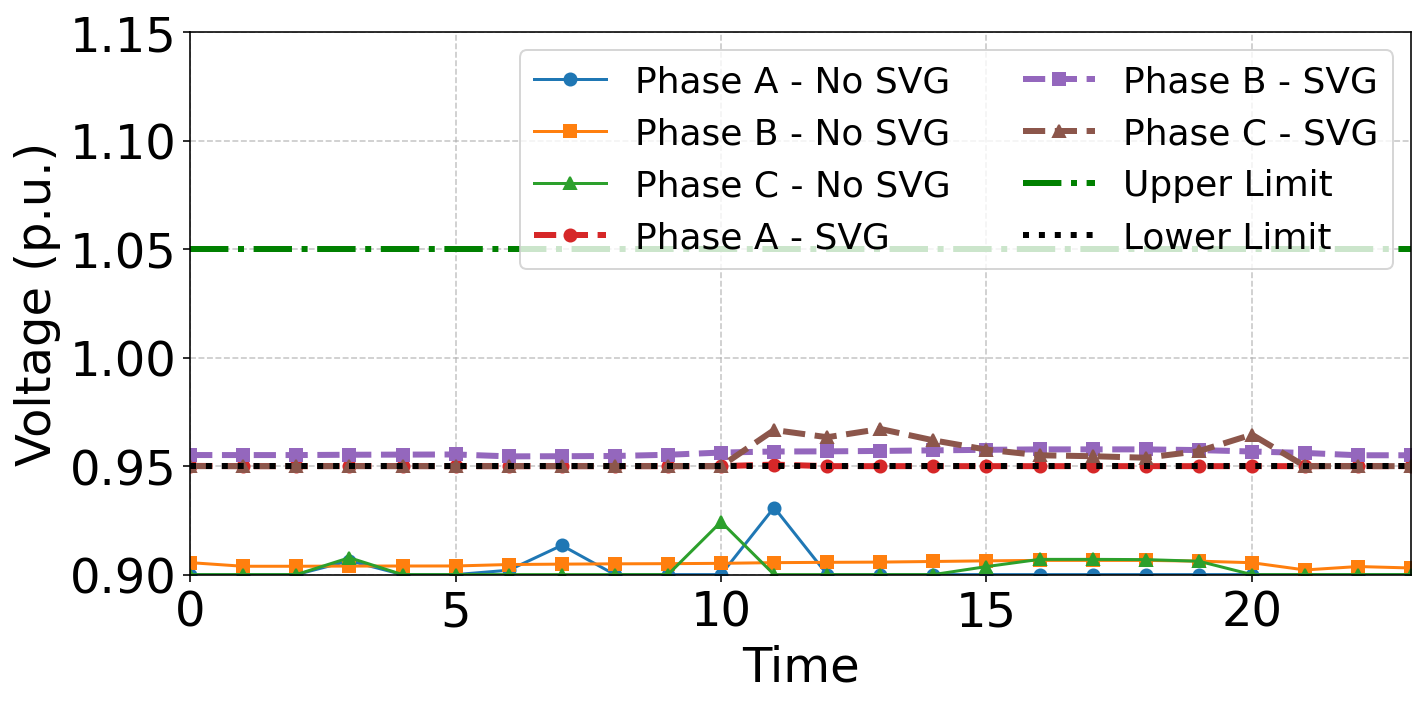}
    \label{fig:voltagevtime741}%
}
\vspace{-0.4cm}
\caption{Voltage at data center nodes for both cases: 1. with $Q^{S}$ ~~2. without $Q^{S}$}
                \label{voltagevtime}

\vspace{-0.7cm}
\end{figure}

\vspace{-0.1cm}
The simulation horizon covers 24 hours. All experiments are implemented in Python 
and solved using the Gurobi optimizer. The proposed framework is evaluated on the IEEE 37-bus distribution test system, with network data obtained from~\cite{Ieee37bus}. Active and reactive loads are generated using uniform distributions $U[100,400]$kW and $U[50,200]$kVAr, respectively. Bus voltage magnitudes are maintained within the range of 0.95–1.05 p.u. on a 4.8 kV base. Four DGs are connected to buses 799, 707, 730, and 736, with their parameters listed in Table~\ref{tab:DG_data}. Data center facilities are located at buses 703, 733, and 741, with power equally distributed across phases. The corresponding data center parameters are adopted from \cite{yang2023distribution}, with a data center power factor of 0.95, and $\tilde{Q}^s_i$ is 350 kVAR for each location. Fig. \ref{DLMPvtime} illustrates the DLMP variation over time for different data center locations and phases. As expected, DLMP values differ across phases in an unbalanced three-phase network. For example, at bus 733 (Fig. \ref{fig:lmpvtime733}) during period 16, the DLMP for phase C is 0.066 \$/kWh, whereas it increases to 0.148 \$/kWh and 0.159 \$/kWh for phases A and B, respectively.
Furthermore, DLMPs vary across network nodes, indicating that data centers can reduce operating costs by selecting more economical locations. For instance, at Fig. \ref{fig:lmpvtime703} during period 14, the DLMP at bus 703 for phase A is 0.108 \$/kWh, while it rises to 0.148 \$/kWh and 0.271 \$/kWh at buses 733 and 741, respectively, as shown in Fig. \ref{fig:lmpvtime733} and Fig. \ref{fig:lmpvtime741}. Fig. \ref{fig:P_dcvtime} illustrates the data center power consumption across all periods for each phase. As observed, higher consumption occurs at bus 703 due to its lower DLMPs. For instance, at period 18, the data center power consumption ($P_{dc}$) reaches 800 kW at bus 703, while it drops to 200 kW and 0.78 kW at buses 733 and 741, respectively. Voltage profiles at the data center buses over all time periods and phases are shown in Fig.~\ref{voltagevtime} for two operating cases: without SVG support $(Q^{s}=0)$ and with SVG-based reactive power compensation $(Q^{s}\neq 0)$. The comparison illustrates the impact of data center reactive power demand on voltage regulation in the unbalanced three-phase feeder. Without SVG support, the increased data center loading causes noticeable voltage degradation, with the most severe drop observed at bus 703 on phase C. As shown in Fig.~\ref{fig:voltagevtime703}, the voltage at this phase drops to 0.926 p.u. during period 15, which is below the acceptable lower limit of 0.95 p.u. After enabling SVG compensation, the voltage at the same bus, phase, and period increases to 0.973 p.u., indicating that local reactive power support effectively mitigates the voltage violation.
Similar improvements are observed at buses 733 and 741, as shown in Fig.~\ref{fig:voltagevtime733} and Fig.~\ref{fig:voltagevtime741}, respectively. The objective value with reactive compensation is \$3708 for $M = 400$. These results demonstrate that SVG support reduces voltage deviations across data center locations and helps maintain voltage magnitudes within the prescribed operating range. 
 Reactive power compensation reduces phase-dependent voltage stress caused by high data center loading in the unbalanced feeder. Future work will study larger systems, sensitivity to power factor, and decomposition methods.

\vspace{-0.4cm}

\vspace{-0.2cm}
\begin{table}
\centering
\caption{DG Parameters}
\begin{tabular}{c c c c c}
\hline
\textbf{DG} & $P^{\max}$ (MW) & $Q^{\max}$ (MVAR) & $a$ ($\$/\text{MW}^2\text{h}$) & $b$ ($\$/\text{MW}\text{h}$) \\
\hline
DG1 & 3 & 1.0 & 0.0045 & 92 \\
DG2 & 3 & 1.5 & 0 & 11 \\
DG3 & 4 & 2.0 & 0 & 7 \\
DG4 & 3 & 1.5 & 0.0035 & 79 \\
\hline
\end{tabular}
\label{tab:DG_data}
\end{table}
\vspace{0.5cm}

\section{Conclusion}
\label{Conclusion}
This paper presented a DLMP-based bilevel optimization framework for coordinating data center operation with an unbalanced three-phase distribution grid. The proposed model captures the interaction between the data center aggregator and the distribution system operator, where the aggregator schedules data center power demand in response to locational prices, while the grid determines DLMPs subject to network constraints. The simulation results on the IEEE 37-bus test feeder showed that the proposed framework can effectively capture phase-dependent DLMP variations, reveal the influence of data center siting on operating cost, and demonstrate the impact of data center loading on bus voltages. In addition, the incorporation of SVG reactive power support was shown to significantly improve voltage profiles and mitigate voltage drops caused by high data center demand. 


\bibliographystyle{IEEEtran}
\bibliography{reference}

\end{document}